# BlackSeaHazNet Scientific Report - EU FP7 IRSES project 2011-2014
## Strachimir Cht. Mavrodiev

Institute Nuclear Research and Nuclear Energy, Bulgarian Academy of Sciences, Sofia,
schtmavr@yahoo.com
**With the help of**
**N. Kilifarska, NIGGG, BAS, Sofia**
**Lazo Pekevski, Skopje Seismological Observatory. Macedonia**
**Georgi Kikuashvili,** *Ivane javakhishvili Tbilisi State University*, **Georgia**

The aims of the project are in the project's title: Complex Research of Earthquake's Forecasting Possibilities, Seismicity and Climate Change Correlations: to create a team for researching the above mentioned problem.

In the next table are presented the number of participated scientists, the Institutes and counties:

| Number | Partner name | Partner short name | Country |
|---|---|---|---|
| 04 | Institute for Nuclear Research and Nuclear Energy, Bulgarian Academy of Sciences | INRNE BAS | Bulgaria |
| 05 | Geophysical Institute, Bulgarian Academy of Sciences | GPHI BAS | Bulgaria |
| 07 | Institute of Oceanography, Hellenic Center for Marine Research | HCMR | Greece |
| 06 | Josef Stefan Institute | JSI | Slovenia |
| 02 | Karst Research Institute, Scientific Research Centre of the Slovenian Academy of Science and Art | SRC SASA SI | Slovenia |
| 04 | Geological Institute, Bulgarian Academy of Sciences | GI BAS | Bulgaria |
| 03 | Central Laboratory of Geodesy, Bulgarian Academy of Sciences | CLG BAS | Bulgaria |
| 03 | Solar-Terrestrial Influences Laboratory, Bulgarian Academy of Sciences | STIL BAS | Bulgaria |
| 03 | "Ss. Cyril and Methodius" University in Skopje | SORM UKIM | Macedonia |
| 05 | Earth and Marine Sciences Institute, TUBITAK Marmara Research Center | TUBITAK MAM | Turkey |
| 06 | Western Survey for Seismic Protection SNCO, Ministry of Emergency Situations | WSSP | Armenia |
| 12 | M. Nodia Institute of Geophysics, Ministry of Education and Science of Georgia | GI MES | Georgia |
| 06 | Seismic Monitoring Center of Ilia State University State University | SMC ISU | Georgia |
| 04 | Institute of Geophysics, National Academy of Sciences of Ukraine | IG NASU | Ukraine |
| 02 | National Antarctic Scientific Centre of the Ministry of Education and Science of Ukraine | NASC | Ukraine |
| 05 | Odessa National Polytechnic University | ONPU | Ukraine |
| 03 | Institute for Nuclear Research, National Academy of Sciences of Ukraine | INR NASU | Ukraine |

The main results achieved are as follows:
1. Creating a group which is able to fulfill a Complex Research of Earthquake's Forecasting Possibilities.

**The main result is statistical prove of imminent forecasting possibility for seismic regional activity on the basis of the geomagnetic monitoring in the framework of special created data acquisition system for archiving, visualization and analysis**.

The approach was developed in the last 10-12 years (Mavrodiev, 2004, Mavrodiev, Pekevski, Jimseladze, 2008, Mavrodiev, Pekevski, Kukiashvili, 2012) and and explanation for everyday monitoring is presenting in http://theo.inrne.bas.bg/~mavrodi

2. Describing of the main result.
2.1. Using data:
The data acquisition system (http://theo.inrne.bas.bg/~mavrodi), applied for **BlackSeaHazNet** every day geomagnetic and earthquake monitoring use:
- the Balkan Intermagnet geomagnetic stations PAG, Bulgaria, SUA, Romania, GCK Serbia, minute data (http://www.intermagnet.org/),
- software for calculation of the daily and minute Earth tide behavior (Dennis Milbert, NASA, http://home.comcast.net/~dmilbert/softs/solid.htm),
- the Earth tide extremes (daily average maximum, minimum and inflexed point) as trigger of earthquakes,
- the data for World A- indices (http:/www.swpc.noaa.gov/alerts/a-index.html).

2.2. The simple mathematics for calculation of the Precursor signal and software for illustration the reliability of forecasting and its statistic estimation (see Fig. 1):
a. The variables $X_m, Y_m, Z_m$ are the component of minute averaged values of Geomagnetic vector or its variations, m=1440.

b. The variables $dX_h, dY_h, DZ_h$ are standard deviation of $X_m, Y_m, Z_m$, calculated for 1 hour (h=1,..,24):

$$X_h = 1/60 \sum_{m=1}^{60} X_m, \text{ where h=1,..,24},$$

$$dX_h = \sqrt{1/60 \sum_{m=1}^{60} \left(1 - \frac{X_m}{X_h}\right)^2}.$$

c. and geomagnetic signal

$$GeomHourSig_h = \sqrt{\frac{dX_h^2 + dY_h^2 + dZ_h^2}{X_h^2 + Y_h^2 + Z_h^2}}$$

d. The A indices are the Low, Medium, High a- indices calculated by NOAA, Space weather prediction center: http://www.swpc.noaa.gov/alerts/a-index.html.

e.  The variable GmSig$_{day}$ is diurnal mean value of GmHourSig$_h$:

$$GeomSig_{day} = 1/24 \sum_{m=1}^{24} GeomHourSig_h$$

and PrecursorSig$_{day}$ is

$$\Pr ecursorSig_{day} = 2 \frac{GeomSig_{day} - GeomSig_{yesterday}}{Amg_{day} + Amg_{yesterday}}$$

f.  The indices of Eq's magnitude value is the distance in hundred km from the monitoring point.

g.  The variable SChtM is the eq's modified energy surface density in the monitoring point [J/km$^2$] :

$$SChtM = \frac{10^{(1.4M+4.8)}}{(40+Depth+Dis\tan ce)^2}$$

h.  The variable PerDayEqSum is the sum of energy density *SChtM* of all eq's, occurred in the time period +/- 2.7 days before and next of the tide extreme. Obviously its value can serve as estimation of regional seismic activity for the time period around tide's extreme.

i.  The variable SumEnergy is the sum of energy density *SChtM* of all eq's, occurred in the day,

j.  The variable TideMinute is the module of Tide vector calculated every 15 minutes,

k.  The variable TideDay is the diurnal mean value in time calculated in analogy of *mass center* formulae

$$Time_{TideDay} = \sum_{m=1}^{360} mTideDay_m / \sum_{m=1}^{360} TideDay_m$$

For seconds and more samples per second the generalization has to calculate geomagnetic components for every minute and correspondingly the GmSig$_{day}$ has to be the mean value for 1440 minutes.

As one can see that the appearance of positive $\Pr ecursorSig_{day}$ value means that in the time period of next Tides extreme ( minimum, maximum or inflex behaviour) the regional seismic activity increases (the bigger value of variable PerDayEqSum).

**So, the described geomagnetic quake approach using monitoring data from nearest geomagnetic station can serve as precursor for imminent estimation of regional seismic activity**.

For boreholes water level data one do not use the A indices data, the GmSig has to be changed with Water level signal WlSig and Precursor signal is only a derivative:

$$\Pr ecursorSig_{day} = WlSig_{day} - WlSig_{yesterday}$$

For statistic estimation of the reliability we calculate the variable day difference

DayDiff = EqTime – TideExtremTime [day]

and calculate its distribution for those earthquakes with biggest values of SChtM.

3. The reliability of regional imminent seismic condition on the basis of Geomagnetic quake approach were tested statistically using the geomagnetic data of monitoring stations:

- Intermagnet PAG (Panagurichte, Bulgaria),

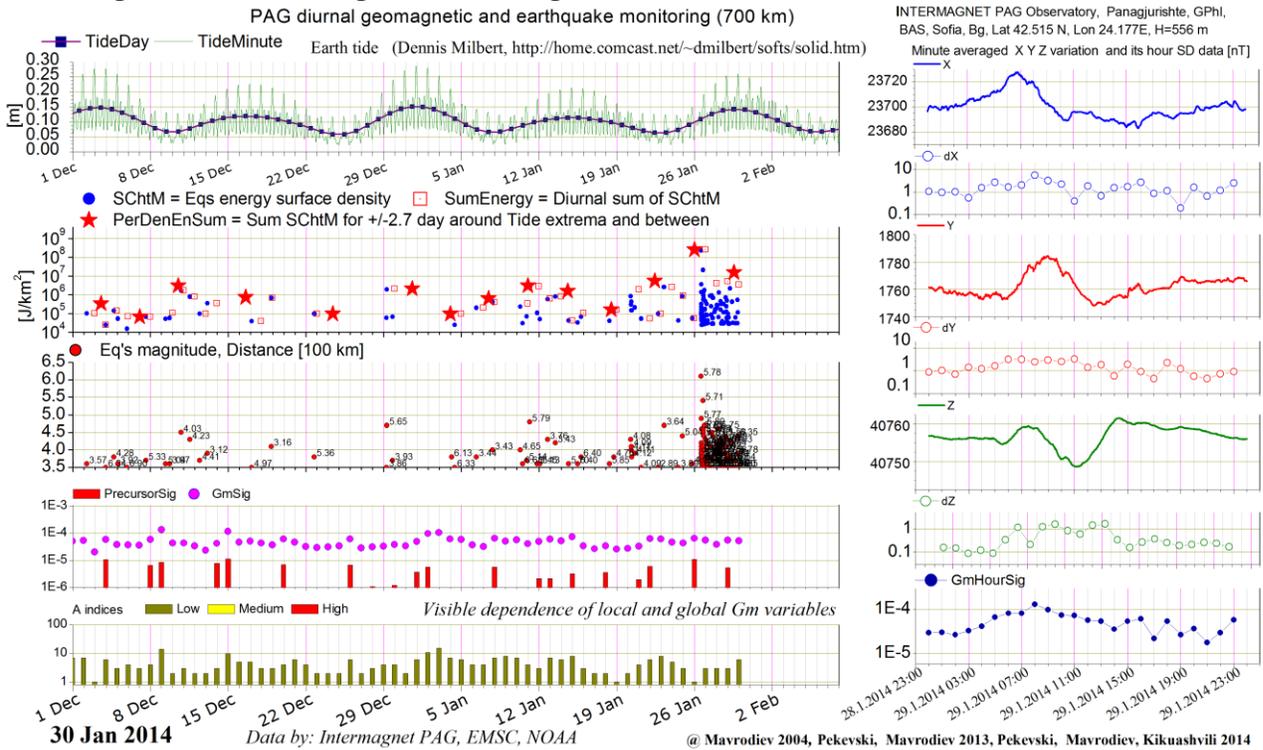

Fig 1. Reliability of the approach for the PAG Intermagnet station located in Bulgaria

For the period from January 1, 2008 to 29 January 2014 were calculated the values of day difference between the times of predicted (by definition we suppose that predicted earthquakes are those which have maximal energy density in the monitoring point and occurred in +/- 2.7 days around tide's extreme) and occurred earthquakes.

As one can see from the next figure the distribution is near to the Gauss one with $hi^2 = 0.89$. The relation between sum of energies of occurred and predicted earthquakes r = 6.48/5.51. **This facts can be interpreted as statistical prove that the geomagnetic quake approach is reliable for estimation of imminent regional seismic conditions.**

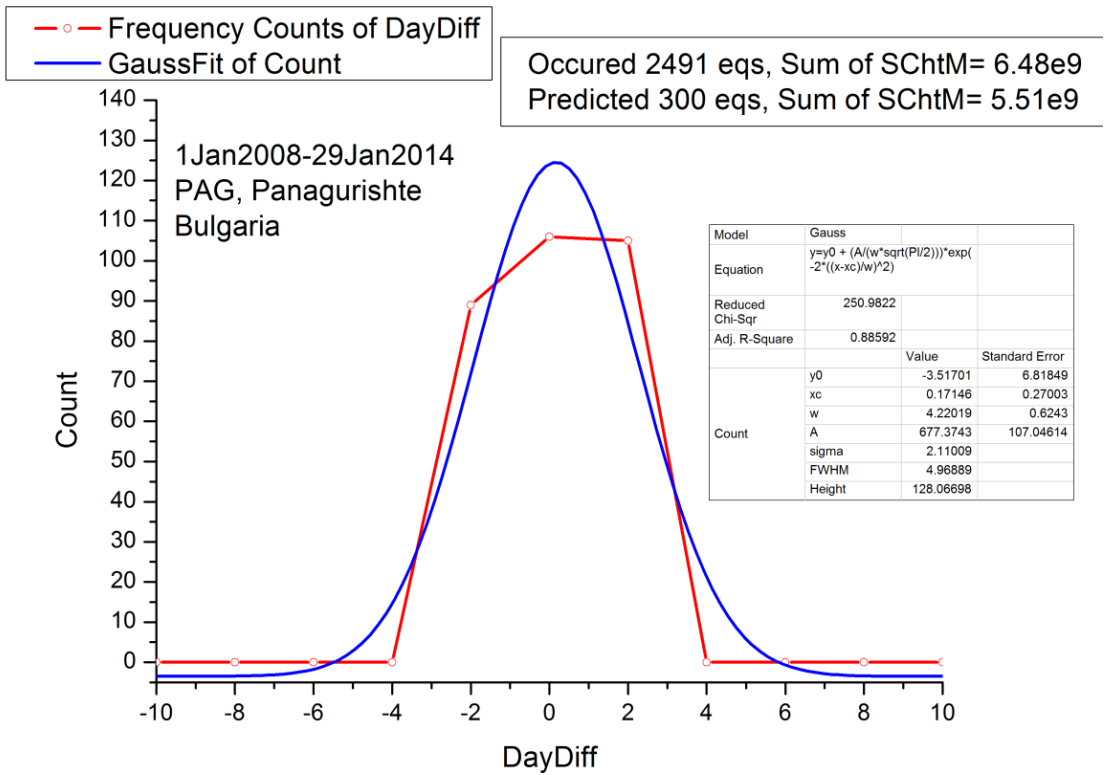

For illustration the map of predicted (column/bar with values logarithm of variable SChtM) and occurred (circles) earthquakes is presented:

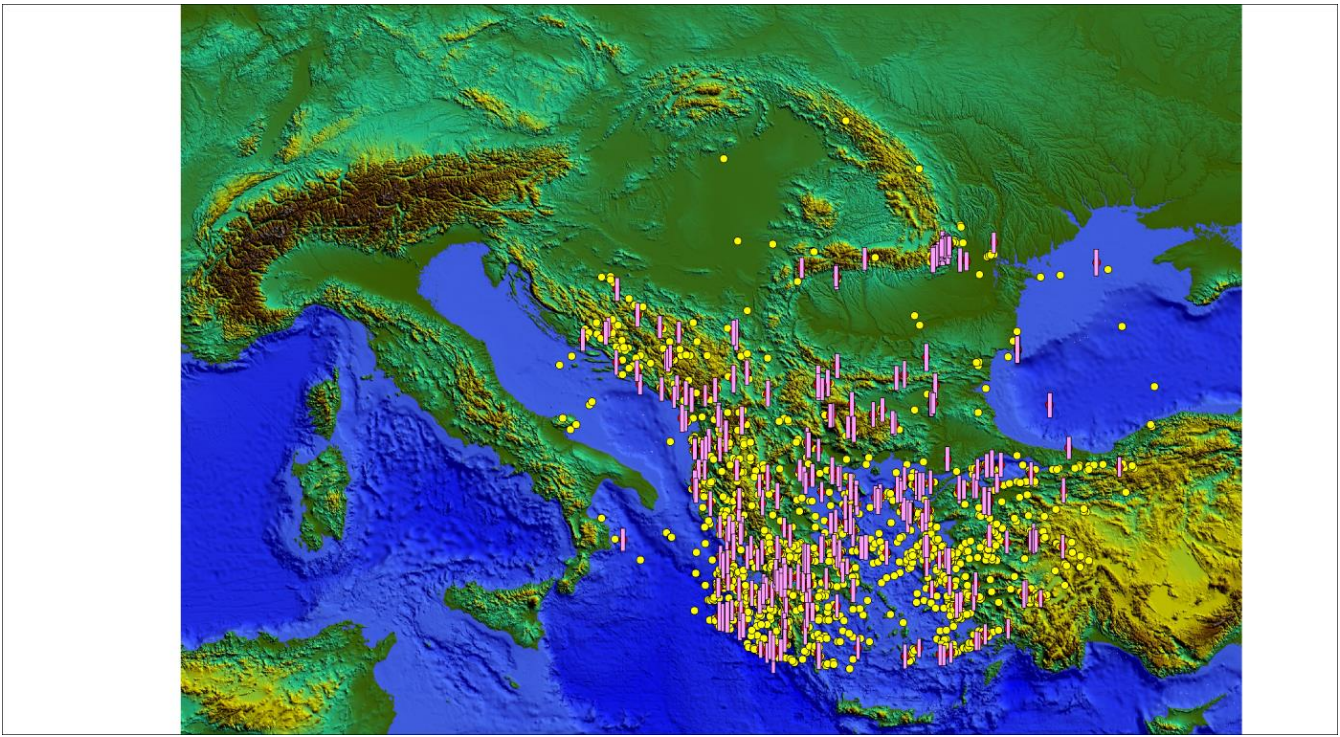

- Intermagnet SUA (Surlari, Romania):

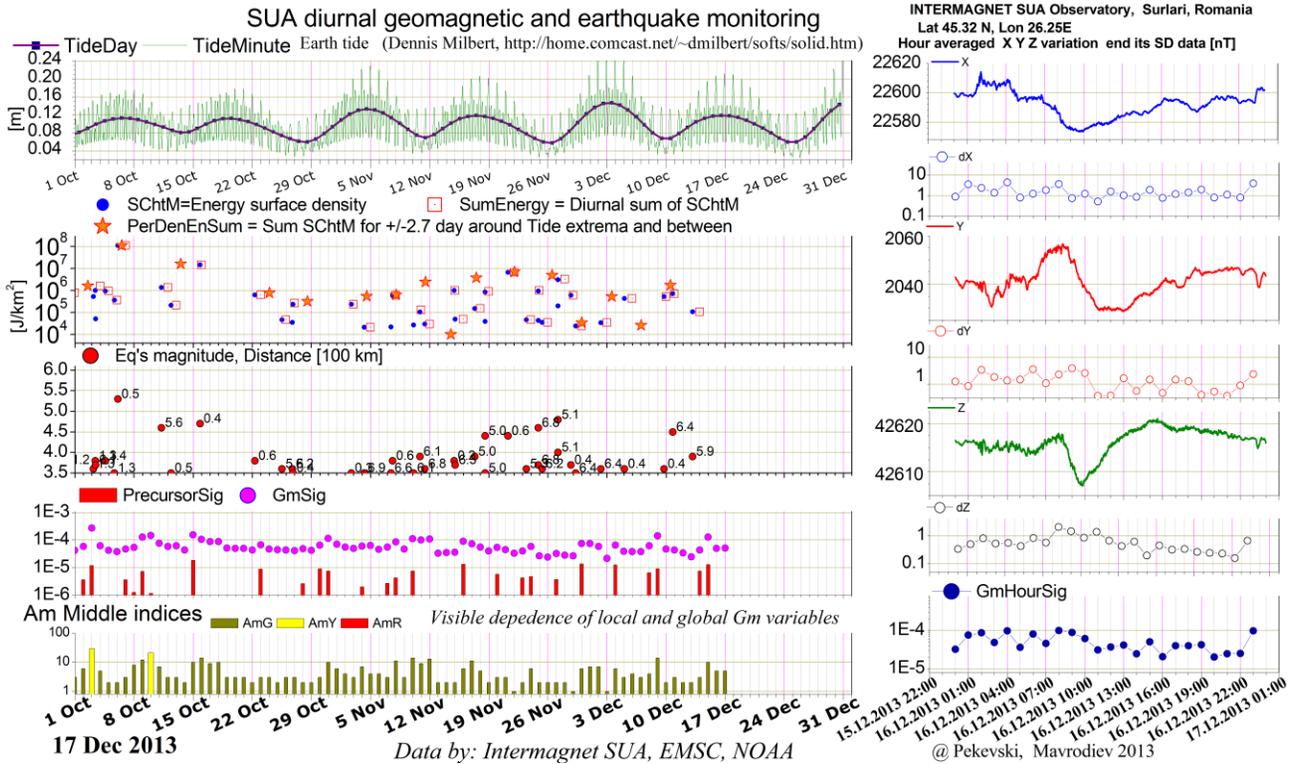

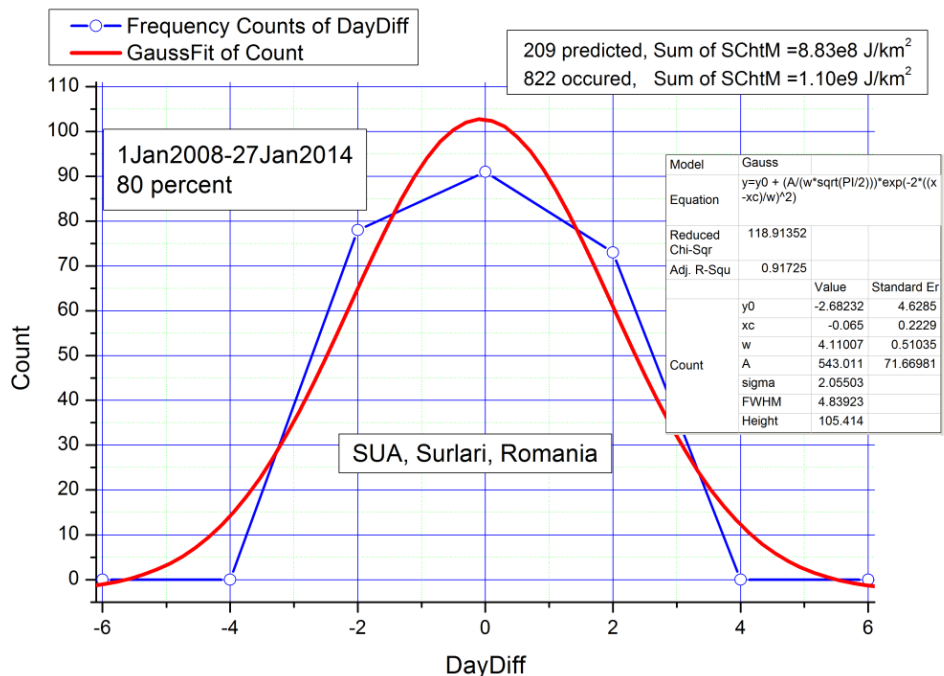

As one can see from the above figure the distribution is near to the Gauss one with hi$^2$ = 0.92. The relation between sum of energies of occurred and predicted earthquakes r = 11.0/8.83

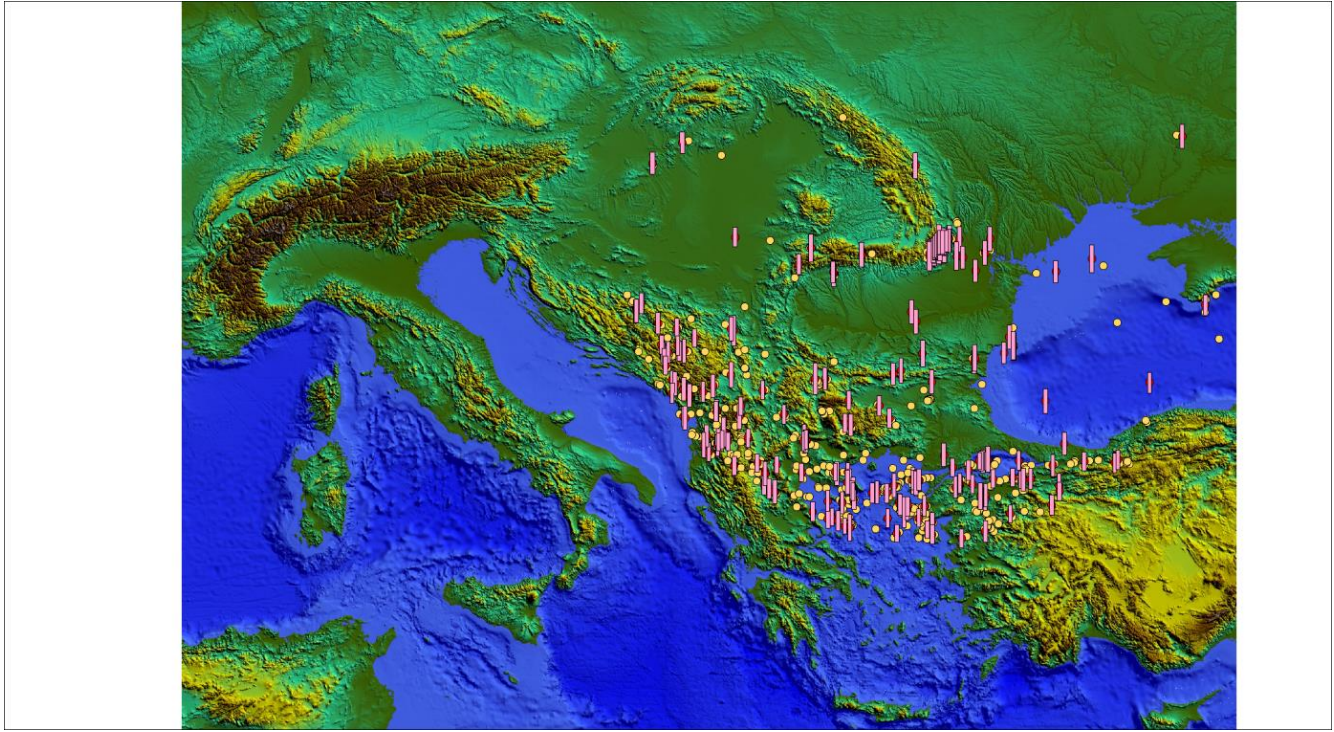

- Intermagnet GCK (Grocka, Serbia)

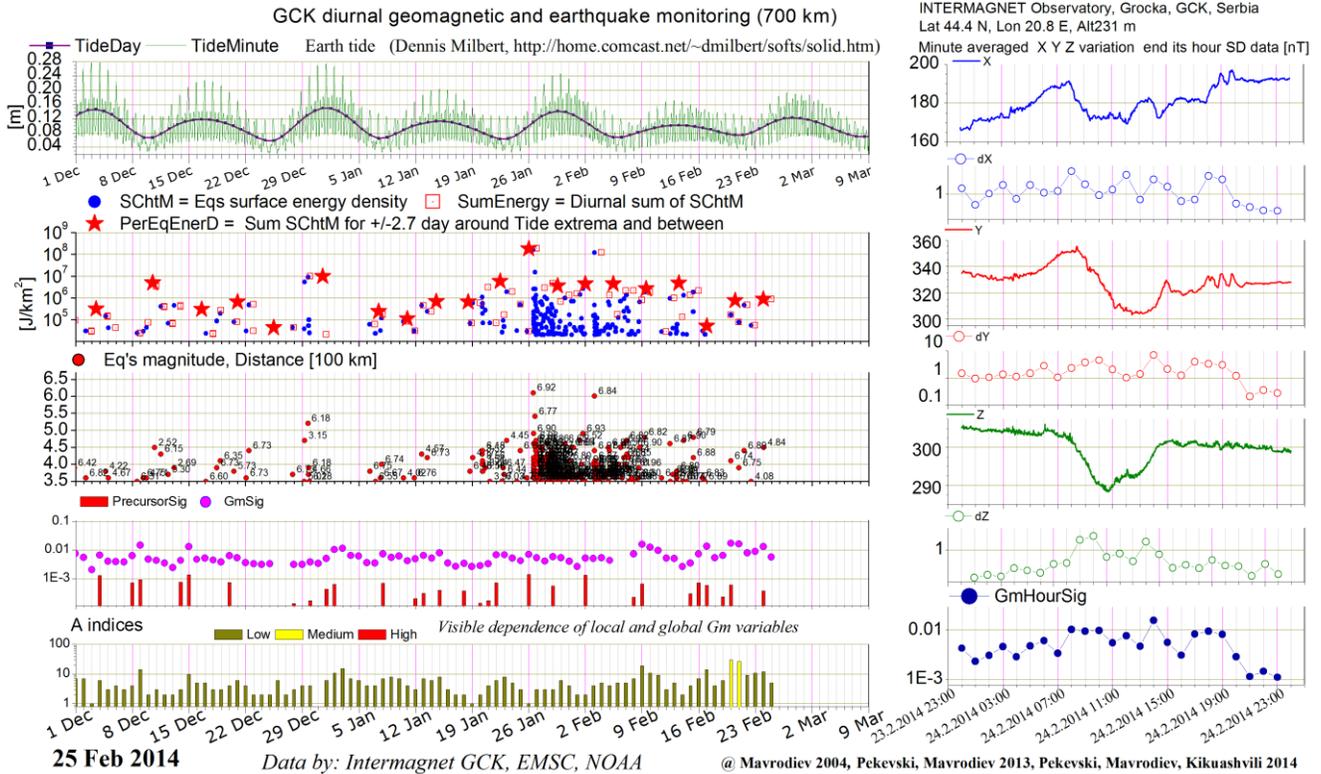

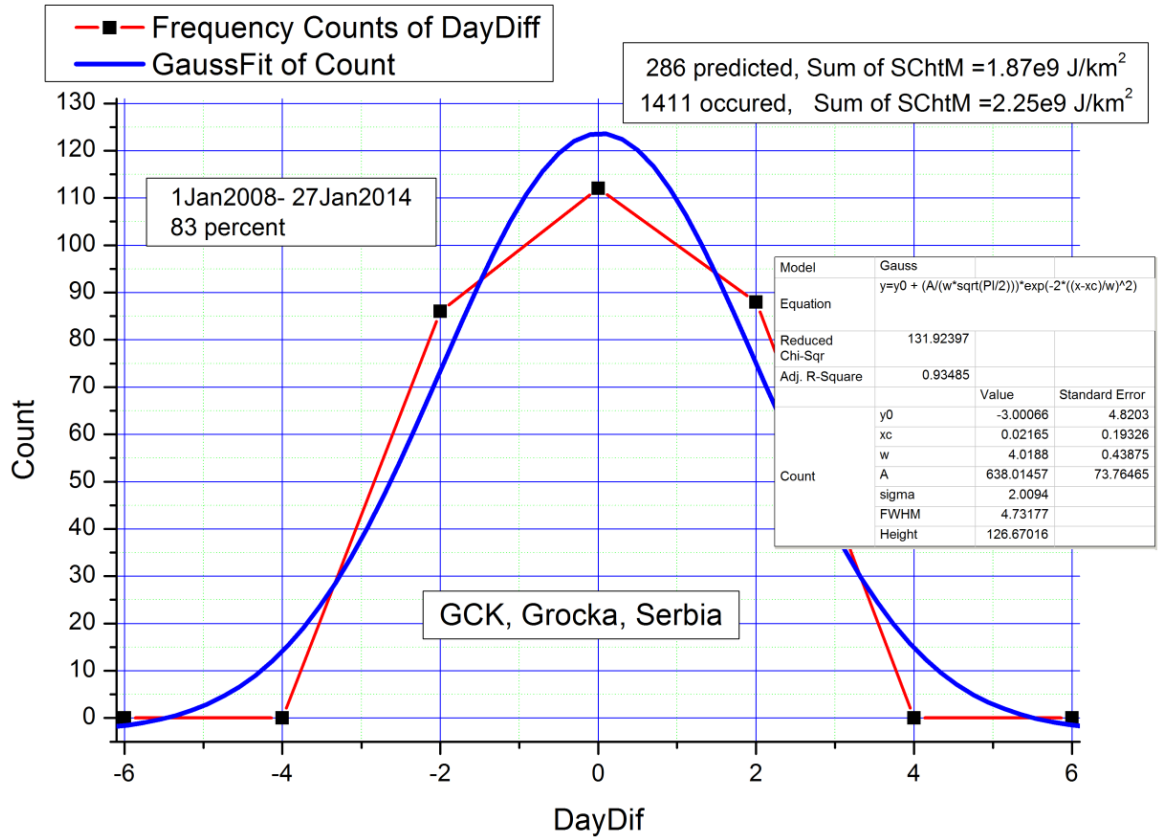

As one can see from the above figure the distribution is near to the Gauss one with hi² = 0.93. The relation between sum of energies of occurred and predicted earthquakes

r = 2.25/1.87

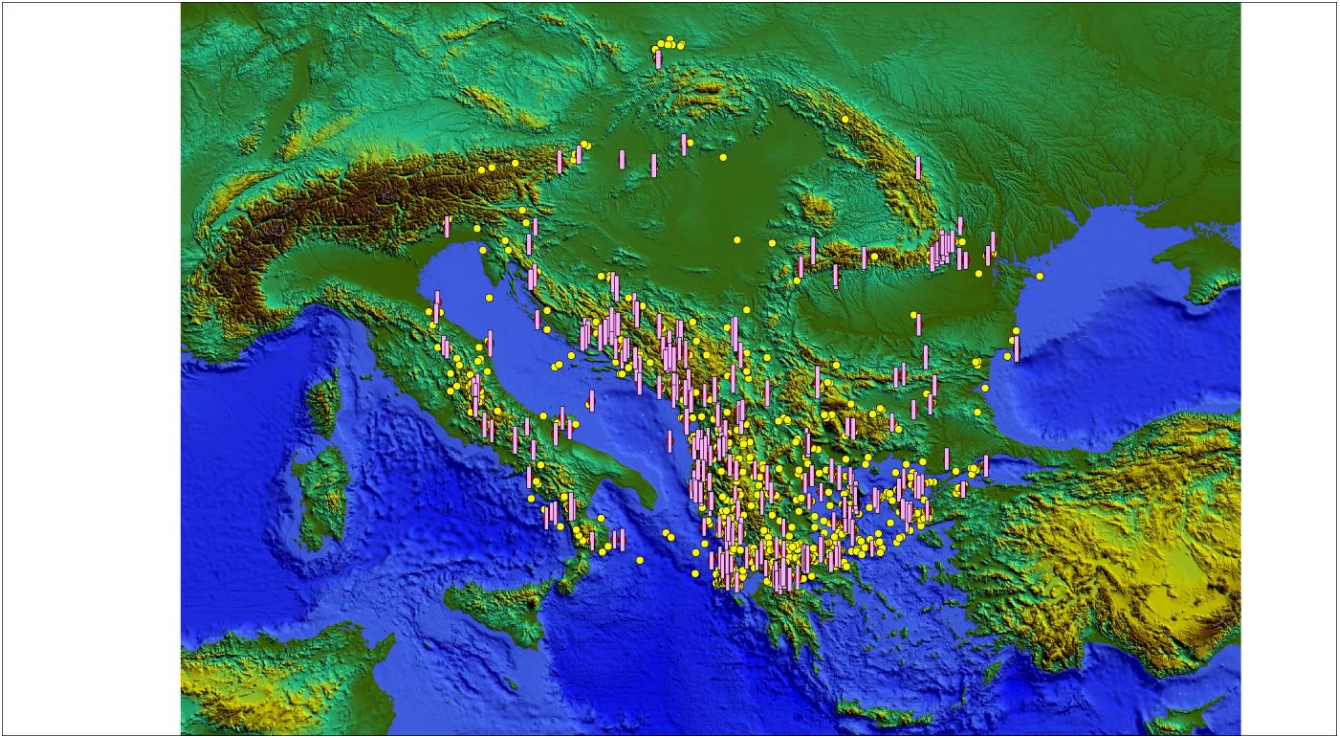

- Intermagnet L'Aquila (AQU, Italy)

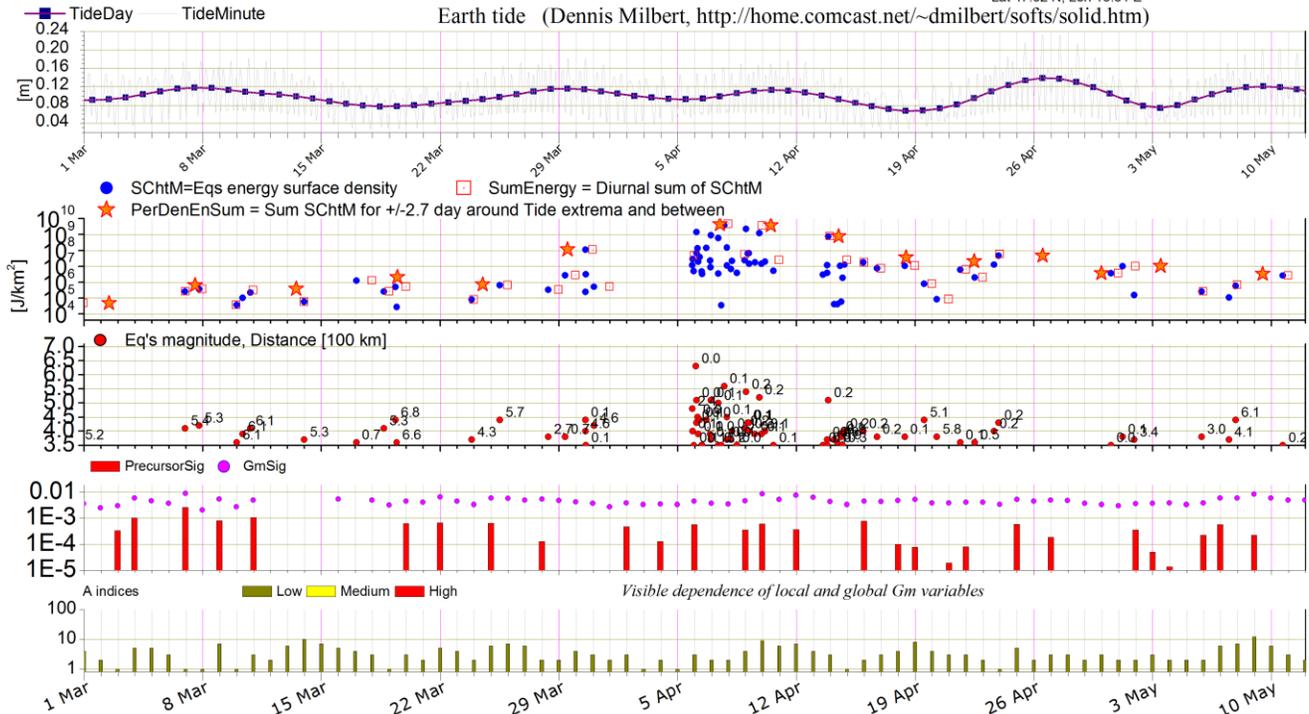

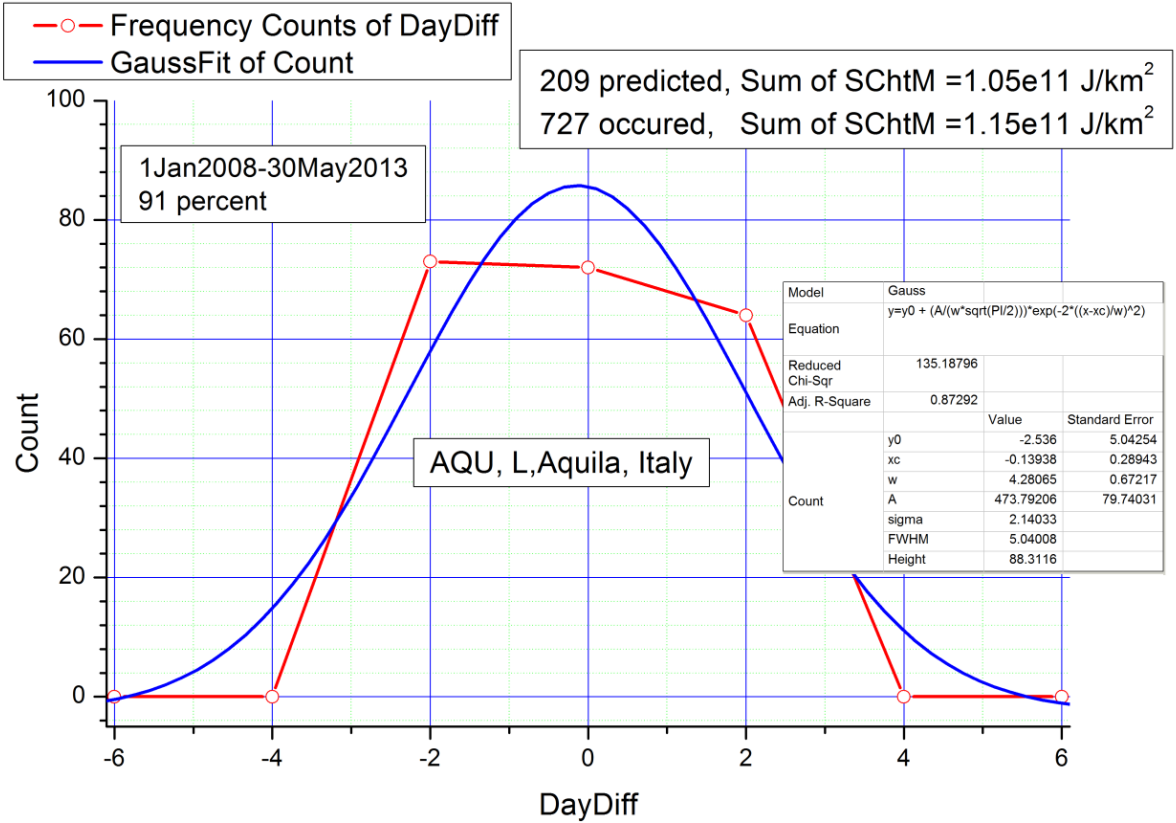

As one can see from the above figure the distribution is near to the Gauss one with hi$^2$ = 0.87. The relation between sum of energies of occurred and predicted earthquakes
r = 1.15/1.05

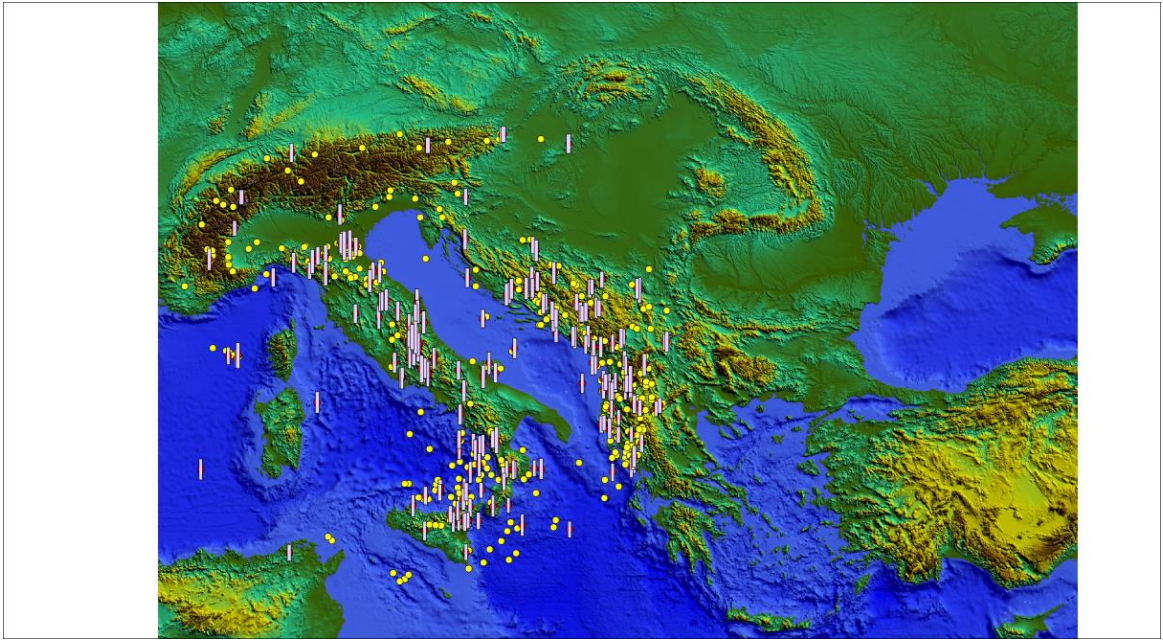

- BlackSeaHazNet second station, Skopje, Macedonia,
- Dusheti, Georgia (fluxgate magnetometers with minutes and second samples),
- MES NSSP Netwok, Armenia, Stepanavan geomagnetic station (Proton magnetometer, hour samples)

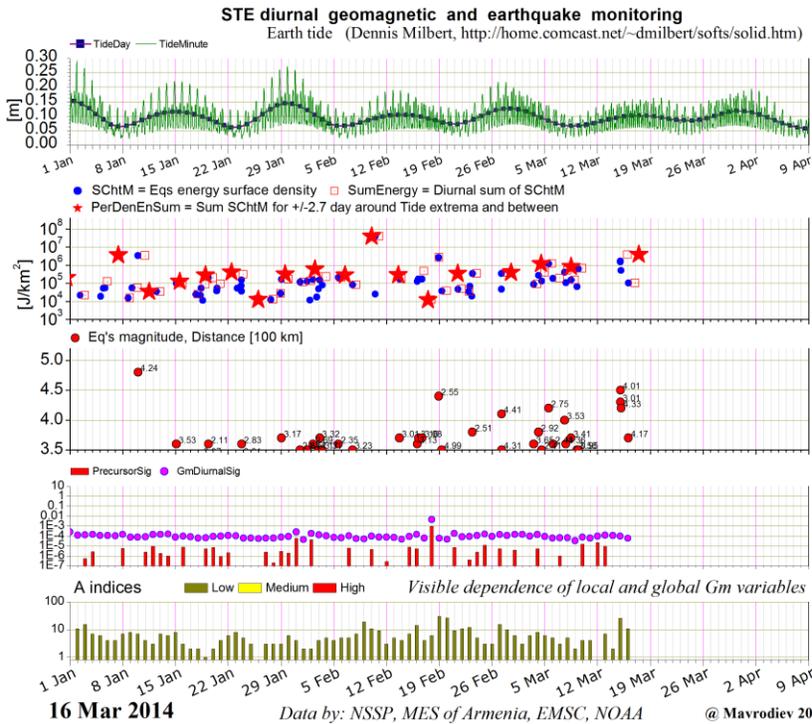
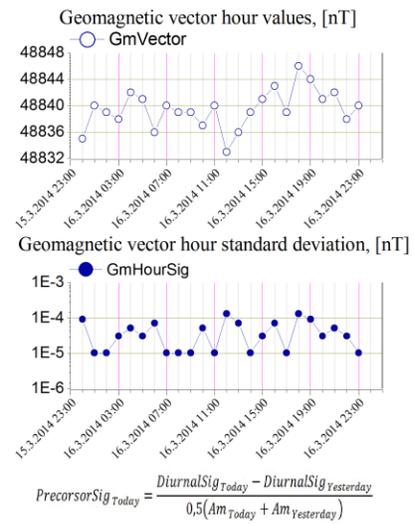

For Kiev and Lvov (Ukraine) data is not possible to do the above conclusions about Gauss distribution, because of irregularity of monitoring and the data analysis.

4. Big world earthquakes and Intermagnet data. During our investigation of relation between regional geomagnetic and seismic activity in areas of interest, close to particular Intermagnet geomagnetic observatories (GMO), it was found that in case of strong earthquake occurred on epicenter distances less than 600 - 1000km from geomagnetic observatory, clear precursor signal was evident.

5. Imminent regional confirmation of forecasting based on the geomagnetic quake (positive jump of PrecursorSig$_{day}$) approach:
- Dusheti, Georgia flux gate second magnetometer - Mw 7.1, depth 7.2 km, 2011, 23 Oct, 36.63N. 43.49E,Van, Turkey earthquake;
- Skopje, Macedonia (second) and Panagurichte (minute) flux gate magnetometers – Mw 5.6, Depth 9.4 km, 42.66 N, 23.01 E, 00.00 hour, 22 May, 2012;
- Grocka, Serbia and Panagurichte (minute) flux gate magnetometers – Mw 6.1, Depth 18km, 26 Jan 2014, 13:55, 38.19 N, 20.41 E;  Mw 6.0, depth 2km, 3 Feb, 2014, 03:08, 38,25 N, 20.32 E, Mw5.6, Depth 9.4km.

6. The acquisition system for archiving, visualization and analysis of the water level variations in boreholes as earthquake precursor was created for Georgia and Armenia data (http://theo.inrne.bas.bg/~mavrodi , http://dspace.nplg.gov.ge/handle/1234/9101):

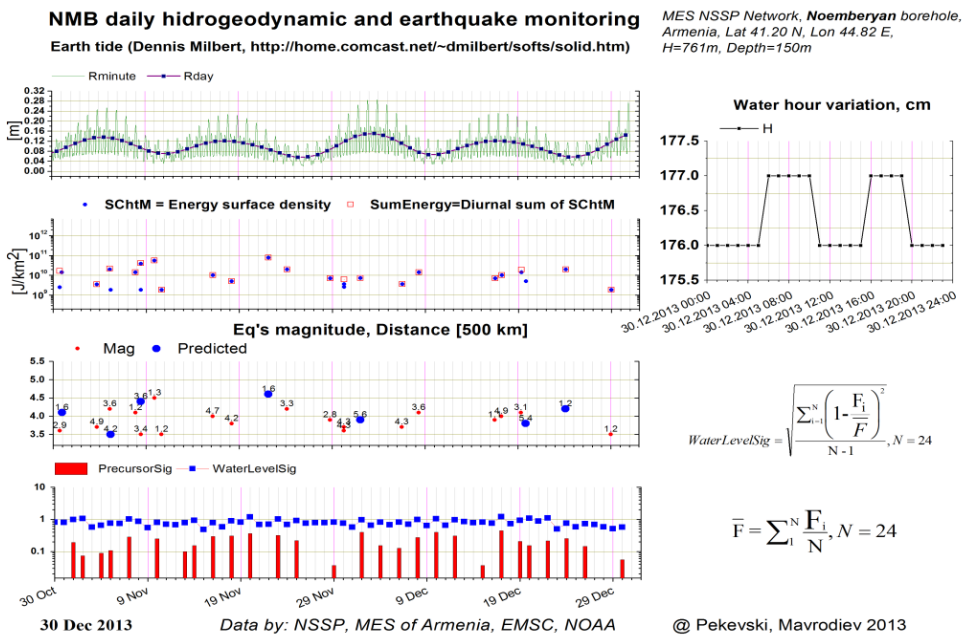

7. Another geophysical network was tested in Ukraine and Antarctica: seismic, meteorological, electromagnetic (VLF), geomagnetic, infrasound, radon monitoring. Estimated probability for earthquakes with M> 5 was 0.8, Vrancha, Romania and earthquakes with M>6.5 was 0.6 for the Scotia Sea region, Antarctica.

8. The reality of Climate Seismicity correlation and axion- geo -nuclear -reactors hypothesis for Climate changes reasons

In the Rusov's talk in Ohrid, Macedonia 2011 workshop was presented the hypothesis and some experimentally argumentations that Solar processes are the host power pacemaker of Earth climate behavior and its seismicity.

The causality link processes are as follows:
- the burn of one Sun axion from two gamma quanta in the field of iron nuclei (the strait Primakoff effect);
- the burn of two gamma quanta (the inverse Primakoff effect) in the field of iron nuclei in the Earth's nuclei and the increasing of temperature, which leads to the activation of geo nuclear set of reactors (Feoktistov type $^{238}$U, $^{232}$Th reactors with fast neutrons) in the canyons on the nuclei's surface;
- as a result there is more heat, more intensive lifting of magma, more activity in the oceans rift zones, more intensive Wegener's plates movement, and, consequently, more seismic and volcanic activity as well as change of climate behavior.
- as well as the discovery of neutrino was based on the conservation laws, we can hope that some estimations for the axions existence, its mass, the spatial distribution and characteristics of geo – reactors will be achieved after creation of the more accurate Earth's heat balance models and the experimental measuring of neutrino's type and energy distribution.

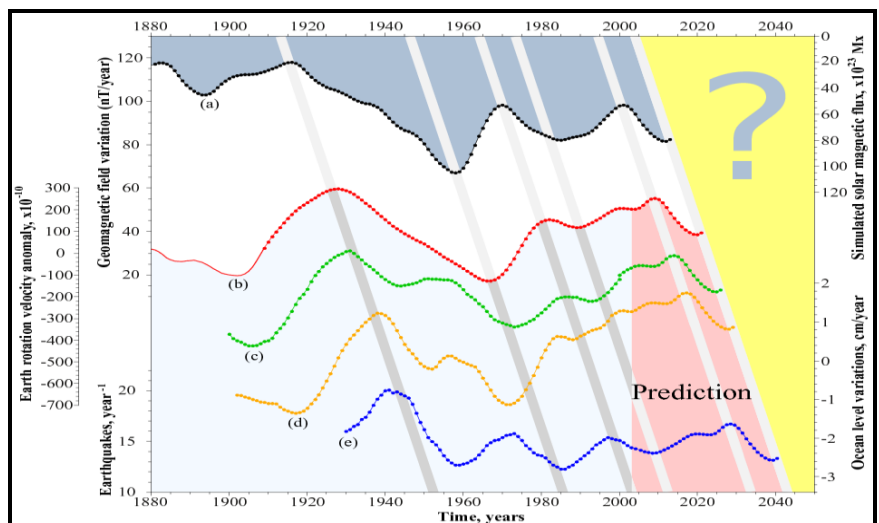

Time evolution: (a) of the variations of magnetic flux in the bottom of the Sun convective zone (tachocline zone); (b) of the geomagnetic field secular variations (Y-component, nT/year), the values of which are obtained at the Eskdalemuir Observatory (England), where the variations (*δY/δt*) are directly proportional to the westward drift of magnetic features; (c) of the variation of the Earth rotation velocity; (d) of the variations of the average global ocean level (PDO+AMO, cm/year); (e) the number of large earthquakes (with magnitude $M > 7.0$). All curves are smoothed by sliding intervals in 5 and 11 years. The pink area is the prediction region. Note: formation of the second peaks on curves (c) – (e) is mainly predetermined by nuclear tests in 1945–1990.

9. Geo electromagnetic measurements: for the first time in Bulgaria territory were measured in the same point the Earth electric currents and geomagnetic field component using the station GEOMAG-02 and magnetometers GEOMAG-02M: Main technical characteristics of MTS GEOMAG-02 are: measuring range of full MF±65000nT; measuring range of MF variations ± 3200nT; resolution of MF variation registration to FLASH-card 0.01nT; temperature drift <0.2nT/°C; tolerance of component non-orthogonally of MF sensor <30ang. min; automatic compensation range of contact MF in each component ±65000nT; EF variation measuring range ±200mV; ±2000mV; resolution of EF variation registration to FLASH-card 1μV; measuring channel frequency band DC - 1 (3, 10)Hz; measuring channel information sampling numbers 10-15 in sec; data averaging during recording to FLASH-card 0.1…60s; capacity of FLASH-card «CompactFlash» (FAT-16, FAT-32)  64MB…64GB; operating temperature range 10°-40°; connecting cable length between MF sensor and electronic unit up to 50m; power consumption 12V; 0.1A.

10. Radon mapping was caring out on the territory Georgia and Slovenia for fixed active fault system and gas concentration monitoring was starting, including in the cave system. Bat there was not enough long time series for receiving the estimation of Radon concentration variations as regional earthquake's precursor.

11. Meteorology and seismicity correlations: Investigation of the possible correlation between meteorology variables and regional seismic activity was started.

12. Ozone and temperature spatial distributions and its possible correlations with regional seismic activity: There is a good correspondence between geomagnetic field, near surface air temperature and pressure spatial distributions in Northern hemisphere during XX century. The alteration of the near tropopause temperature (by $O_3$ variations at these levels) changes the amount of the water vapor in the driest part of the upper troposphere/lower stratosphere. Application of non-linear statistical methods for analysis of climatic and magnetic field data reveals the important role of energetic particles and lower stratospheric ozone in climate variations.

13. Electromagnetic scanning: The research of deep Earth's crust structure and upper mantle study using the inverse problem analysis of the Earth electromagnetic radiation in radio diapason, measured with "Astrogon" device was performed in Greece and Bulgaria in different profiles. The device is a passive sensor type sensitive to the three components of the magnetic field within a wide range of frequencies (5 – 100 kHz). The comparison with geological knowledge for the Crust in the profiles and the inverse problem results give a hope for perspectives of such kind of studies and that the method has to be included in the permanent regional monitoring. Really, during the project fulfilling, the method of the electromagnetic tomography of the Earth crust and upper mantle was developed, which gives the possibility to determine the location of the future earthquake hypocenters as the most stressed volume of the crust (or mantle). Moreover, the stationary EM measurements by the same device (or new one DS-4, designed and produced during the project fulfilling) show the existence of electromagnetic precursor in a wide frequency diapason, which coincide with that of low frequency signal in Intermagnet data. So the base of the project task solution – where and when – is grounded.

14. Other precursors research are as follows: The first is TM 71 extensometer monitoring, which is carried out to observe micro-displacements along fault movements (or landslide movements connected with active tectonics) in karts caves, fault scarps or in trenches where was found anomaly in velocities 3-4 months before regional strong earthquake. The second represents 2D displacements of static vertical pendulum in 25m deep karst shaft that are registered each 10 seconds. Changes in stress direction are detected. The studies were oriented towards the aim to connect the periods of micro-displacements with local and worldwide seismicity. The third represents the temperature monitoring of two sulphidic waters, which are situated near important regional faults. The fourth includes microbiological monitoring site on the fault planes in the Postojna Cave (Slovenia) to find the possible connection between microbial biomass and tectonic displacements.

15. Schuman resonance measurement device: it was developed the construction design and software for measuring device were tested.

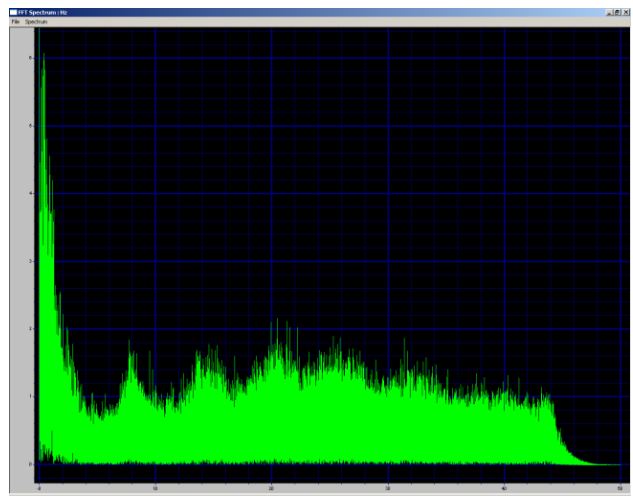

16. The website of the Project was created:
http://theo.inrne.bas.bg/~mavrodi/blackseahaznet/

During the project many young scientists visited research centers of the Black Sea region, which facilitated their contacts with colleagues. They take part in project conferences, seminars, joint field works and processing data in hosting countries. As a result, joint publications were published in journals and conference proceedings.

In the case of project prolongation until the end of 2014 in the frame of the remaining budget (around 34% or approximately 160000EUR) **the main expected results which are based on the project current achievements will be:**

A. Preparation of the project proposal for regional electromagnetic monitoring under, on and above Earth's surface and near space and as well meteorological data for creating of complex data acquisition system on the basis of which to start solving the inverse problem for regional imminent forecasting of time, coordinates, depth, magnitude and intensity of incoming earthquakes.

B. Creation of project proposal/s for Sun – Earth interaction balances models which describe its influences on climate change, seismicity, volcanism and continental plate's movement.

C. Development of long term collaboration for complex research in the framework of bilateral, regional and other European 2020 programs.

**Conferences presentations , papers**
The members of Project participated 8 International Conferences, Congresses and Workshops with presentations as well as published many papers.
1. Kilifarska N.A., Bakhmutov V.G., Martazinova V.F., Melnyk G.V., Ivanova E.K., Geomagnetic field influence on the climatic parameters, Second Intern. Conference on "Актуальные проблемы электомагнитных зондирующих систем", Kiev, Ukraine, 1-4 October, 2012.
2. Kilifarska N.A., Climate Variability Initiated by Helio- and Geomagnetic fields−Evidences and Mechanisms, TOSCA MS meeting, 29 Sept−4 Oct, 2013, Prague.
3. Kilifarska N.A., Near tropopause O3 – variability and climate implication, 12$^{th}$ Scientific Assembly of the IAGA, August 25-31, 2013, Mérida, México.
4. Kilifarska N., Bakhmutov V., Melnik G., Atmospheric ozone and Antarctic climate, VI International Antarctic Conference, May 15–17, 2013, Kiev, Ukraine
5. Kilifarska N., Bakhmutov V., Melnik G., Geomagnetic field as a driver of Climate variability, XII Intern. Conf. on Geoinformatics: theoretical and applied aspects, 13-16 May 2013, Kiev, Ukraine

**Acknowledgments:**

**We are very thankful to all participation of the project for active work and enthusiasm.**

**Without the everlasting efforts of coordinators Natalia Kilifarska, Christos Tsabaris, Jania Vaupotich, Stanka Sebela, Nikolai Dobrev, Katia Georgieva, Lazo Pekevski, Erham Alparslan, Hrachya Petrosyan, Tamaz Chelidze, George Melikadze, Vladymyr Bakhmutov, Vazira Martazinova, Oleksandr Lyashchuk, Vitalii Rusov and Volodymyr Pavlovych the results of Project could not be so good.**

**We have to note the special roll and big work of Person in charge of administrative, legal and financial aspects of the Project B.Vachev.**

We would like to thank heartily the REA project officers heartily Dr. Oscar Perez-Punzano and Dr. Atantza Uriarte-Iraola, for their inavaluable support in the process of negotiation and executing the project.

**We would like also to express our sincere gratitude to the DG Institute for Nuclear Research and Nuclear Energy, BAS in the face of Directore Assoc. Prof. Dr.**

**Dimitar Tonev and Vice Director Assoc. Prof. Dr. Lachezar Georgiev, for the encouragement and valuable assistance, as well as to Dr. Frank Marx, and in his face – to the European Commission, for the financial support for the BlackSeaHazNet project realization.**